\documentclass[letter]{ptptex}

\usepackage{graphicx}
\newcommand{\VEV}[1]{\langle{#1}\rangle}
\newcommand{\chibar}{{\bar{\chi}}}
\newcommand{\Psfig}[2]{\includegraphics[width=#1]{#2}}
\newcommand{\Feff}[1]{{\cal F}_\mathrm{#1}}
\newcommand{\Vq}{{\cal V}_\mathrm{q}}

\newcommand{\Od}{{\cal O}}
\newcommand{\Vp}{V^{+}}
\newcommand{\Vm}{V^{-}}

\newcommand{\Kern}{K}
\newcommand{\comment}[1]{}
\title{%        %You can use \\ for explicit line-break
Quarkyonic matter in lattice QCD at strong coupling
}
\author{
Kohtaroh \textsc{Miura}$^1$,
Takashi Z. \textsc{Nakano}$^2$ 
and
Akira \textsc{Ohnishi}$^1$
}
\inst{%     %Affiliation, neglected when [addenda] or [errata]
$^1$Yukawa Institute of Theoretical Physics, Kyoto University,
Kyoto 606-8502, Japan\\
$^2$Department of Physics, Kyoto University, Sakyo-ku, Kyoto 606-8502
}

\abst{%         %This abstract is neglected when [addenda] or [errata].
We study the phase diagram of quark matter
at finite temperature and density
in the strong coupling lattice QCD
with one species of unrooted staggered fermions
including finite coupling ($1/g^2$) effects 
for color SU($N_c$).
We find that we may have partially chiral restored medium density matter
at $N_c=3$,
which would correspond to the quarkyonic matter suggested at large $N_c$.
}

\begin{document}

\maketitle

%\section{Introduction}\label{Sec:Introduction}

The QCD phase transition at finite temperature ($T$)
is the latest vacuum phase transition in our universe,
and it can be experimentally investigated at RHIC and LHC.
Phase transitions at finite chemical potential ($\mu$) in dense matter
may be also realized during black hole formation or in neutron stars,
where we have the following central question.
{\em ``What is the next to the hadronic Nambu-Goldstone phase
in the larger $\mu$ direction ?''}
Monte-Carlo (MC) simulations are not yet reliable 
in the region $\mu/T>1$ because of the notorious sign problem,
then it is necessary to invoke some approximations
such as the large number of colors ($N_c$)~\cite{QY}
or the strong coupling limit (SCL)~\cite{SCL,DKS,KMOO,Jolicoeur:1983tz,FP,Bilic,Ohnishi2007,Bringoltz}
in order to answer this question in QCD.

Recently, McLerran and Pisarski have shown that
the {\em next} phase at large $N_c$ should be
the so called {\em quarkyonic} phase,
in which the colors are confined and the baryon density is high~\cite{QY}.
At large $N_c$, gluon contribution to the pressure $\sim \Od(N_c^2)$
is larger than those
from hadrons $\sim \Od(1)$ and quarks $\sim \Od(N_c)$,
then the deconfinement transition temperature $T_d$ is independent
of the quark chemical potential $\mu$ 
as far as it is moderate $\mu \sim \Od(1)$.
In the confined region $T < T_d$,
quark number density is exponentially suppressed 
if $\mu$ is below the quark mass $m_q \sim \Od(1)$,
but it rapidly grows at $\mu>m_q$ and soon reaches high density $\sim \Od(N_c)$.
This dense matter has a characteristic feature that
it is made of {\it quark}s,
while only bar{\it yonic} excitations are allowed
because it is confined.
If the quarkyonic phase is the {\em next} at $N_c=3$,
it may be formed at high densities
in compact astrophysical phenomena
or in heavy-ion collisions at 10-100 $A$ GeV.
MC results with the density of state method
also show the transition to high density phase~\cite{FKS},
but its nature is not yet known.
Very recently, quarkyonic matter is found to exist at $N_c=3$
in effective models of QCD~\cite{Glozman,PNJL},
while the results may depend on the details~\cite{McLerran:2008ua}.
It is now very important and urgently required to discuss
the possibility of the quarkyonic matter phase in QCD for $N_c=3$.

%%%%%%%%%%%%%%%%%%%%%%%%%%%%%%%%%%%%%%%%%%%%%%%%%%%%%%%%%%%%%%%%%%%%%%%%
\begin{figure}[bt]
\centerline{~\Psfig{7.0cm}{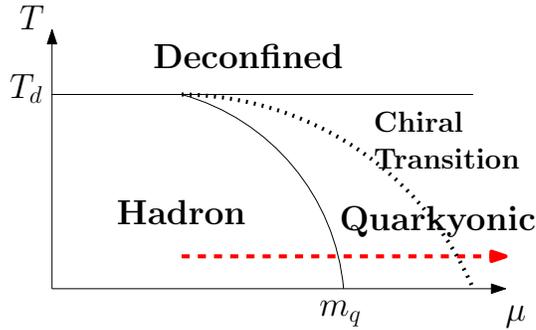}~}
\caption{
Schematic %picture of the 
phase diagram with the quarkyonic matter.
}
\label{Fig:QY}
\end{figure}
%%%%%%%%%%%%%%%%%%%%%%%%%%%%%%%%%%%%%%%%%%%%%%%%%%%%%%%%%%%%%%%%%%%%%%%%

The above discussion tells us that 
the quarkyonic transition, the transition from the 
chiral broken Nambu-Goldstone (NG) phase to the quarkyonic matter,
is characterized by the quark number density.
If the quarkyonic matter exists, 
the quarkyonic transition occurs at $\mu\sim m_q$
where density rapidly grows as $\rho_q=\mathcal{O}(1)\to\mathcal{O}(N_c)$, 
and the chiral restoration follows at higher chemical potential
as shown in Fig.~\ref{Fig:QY}.
In this paper, we discuss the possibility of the quarkyonic matter
in the strong coupling lattice QCD
(SC-LQCD)~\cite{SCL,DKS,KMOO,Jolicoeur:1983tz,FP,Bilic,Ohnishi2007},
which is another powerful tool in studying dense matter.
We take account of the finite coupling effects
in the next-to-leading order (NLO, $1/g^2$)
~\cite{Jolicoeur:1983tz,FP,Bilic,Ohnishi2007},
and introduce an auxiliary field representing the quark number density $\rho_q$
as an order parameter in addition to the chiral condensate $\sigma$.
In a previous work on the phase diagram with NLO effects~\cite{Bilic},
the order parameter representing $\rho_q$ was not introduced.
As shown later, the multi-order parameter ($\sigma$ and $\rho_q$) treatment
is essential in understanding the quarkyonic transition.
In SC-LQCD, we consider the situation where
$g^2$ is large while $N_c$ is fixed.
This condition is somewhat different from that assumed in the large $N_c$
argument, where $N_c$ is assumed to be large and the 't Hooft coupling
$N_cg^2$ is fixed.
As we show later, we find the quarkyonic matter to appear
in some region of $1/g^2$ at $N_c=3$.
This observation together with the large $N_c$ discussion~\cite{QY}
support the existence of the quarkyonic matter
in a wide region of the $(N_c, 1/g^2)$ plane.

We start from the lattice action
with one species (unrooted, four flavors) of staggered fermions ($\chi$)
in the lattice unit,
\begin{align}
S_\mathrm{LQCD}
=\frac12\sum_x \left[ V^+(\mu) - V^-(\mu)\right]
+\sum_x m_0 M_x
+S^{(s)}
-\frac{1}{g^2}\sum_\square
	\mathrm{tr} \left[ U_\square + U^\dagger_\square \right]
\ ,\label{Eq:SLQCD}
\end{align}
where $m_0$ and $\mu$ are 
the bare quark mass and the lattice chemical potential, respectively.
The mesonic composites are defined as
$M_x=\chibar_x\chi_x$,
$\Vp_x=e^{\mu}\chibar_xU_{0,x}\chi_{x+\hat{0}}$
and
$\Vm_x=e^{-\mu}\chibar_{x+\hat{0}}U_{0,x}^{\dagger}\chi_x$,
where sum over color indices are assumed.
The spatial hopping action of quarks $S^{(s)}$
is given as,
\begin{align}
S^{(s)}
=&\frac{1}{2}\sum_{x}\sum_{j=1}^d\eta_{j,x} \left[
\chibar_x U_{j,x} \chi_{x+\hat{j}}
-\chibar_{x+\hat{j}} U^\dagger_{j,x}\chi_x
\right]
\ ,
\end{align}
where
$d$ is the spatial dimension
and $\eta_{j,x}=(-1)^{x_0+\cdots+x_{j-1}}$ is the staggered sign factor.
The gluon degrees of freedom are described by
the temporal and spatial link variables and plaquettes,
$U_{0}$, $U_{j}$, $U_\square$.
This lattice action is invariant under the chiral transformation
$\chi\to e^{i\theta\epsilon_x}\chi$
with the $\gamma_5$-related factor $\epsilon_x=(-1)^{x_0+\cdots+x_{d}}$.

In the strong coupling region ($g\gg 1$),
we can evaluate the plaquette effects
through the expansion in the power series of $1/g^2$
(strong coupling expansion).
In each order of $1/g^2$ or the number of plaquette, 
we can exactly carry out the integral over link variables
by using the SU($N_c$) group integral formulae,
$\int dU~ U_{ab} U^\dagger_{cd} = \delta_{ad}\delta_{bc}/N_c$
and so on,
and obtain the effective action of hadronic
composites~\cite{SCL}.

\begin{figure}
\centerline{~\Psfig{8.5cm}{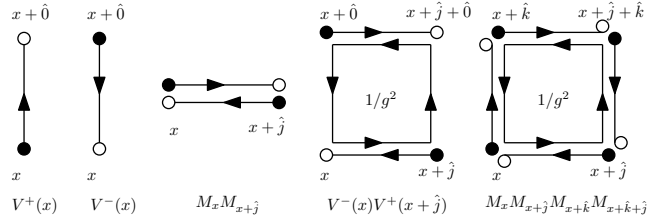}~}
\caption{
Effective action terms
in the strong coupling limit and NLO ($1/g^2$) corrections.
Open circles, Filled circles, and arrows show $\chi$, $\chibar$, and
$U_\nu$, respectively.
}
\label{Fig:plaq}
\end{figure}

%...[ SCL ---------------------------------------------------------------------%
Before discussing the NLO effects,
we briefly summarize the procedure to obtain the effective potential
in the leading order, i.e. the strong coupling limit (SCL),
where the sophisticated framework based on
the $1/d$ expansion~\cite{largeD},
mean field approximation
and finite $T$ treatments of the quark determinant~\cite{DKS,FP},
has been established.
After the spatial link ($U_j$) integral,
we obtain the hadronic hopping action
shown in the third graph in Fig. \ref{Fig:plaq}~\cite{SCL}
from the spatial quark action $S^{(s)}$.
The SCL effective action is given as,
\begin{align}
S_\mathrm{SCL}
= \frac12\sum_x\left[V^+(\mu)-V^-(\mu)\right]+\sum_x m_0 M_x
%\nonumber\\
-\frac{b_\sigma}{2d}\sum_{x,j>0}M_xM_{x+\hat{j}}
+{\cal O}\Bigl(\frac{1}{\sqrt{d}},\frac{1}{g^2}\Bigr)
\ ,
\label{Eq:SCL_Action}
\end{align}
where $b_\sigma=d/2N_c$.
In this work, we adopt the leading order terms of
the $1/d$ expansion~\cite{largeD}
and omit higher order terms ${\cal O}(1/\sqrt{d})$.
In the $1/d$ expansion,
the leading term $\sum_j M_xM_{x+\hat{j}}$ is assumed to remain finite
at large $d$,
then the quark fields scale as $\chi,\bar{\chi} \propto d^{-1/4}$
and higher power terms of quarks are found to be suppressed
as ${\cal O}(1/\sqrt{d})$ for $N_c \geq 3$.
Through the Hubbard-Stratonovich (HS) transformation,
the chirally invariant four-fermi term $S_\mathrm{SCL}$ is converted 
to the quark mass term $b_\sigma\sigma\chibar\chi$,
where the finite chiral condensate $\sigma=-\langle M_x\rangle$ 
breaks the chiral symmetry spontaneously
and generates the quark mass dynamically.
We can carry out the Gauss integral over the quark fields $\chi$ and $\chibar$
in the so-called finite temperature treatment:
The determinant of the temporal quark hopping matrix 
with the anti-periodic boundary condition is obtained
by utilizing the Matsubara product technique~\cite{DKS}
or the recursion relation~\cite{FP}.
It is possible to carry out the temporal link integral of this determinant
over $U_0$ in the Polyakov \cite{DKS} or temporal gauge~\cite{FP}.
Then the effective potential is obtained as,
\begin{align}
\Feff{eff}^{\mathrm{SCL}}
&=\frac{b_{\sigma}}{2}\sigma^2+{\cal V}_q(m_q;\mu,T)\ ,\\
{\cal V}_q(m_q;\mu,T)
&=-T\log\left[
	X_{N_c}
	+2\cosh\Bigl[\frac{N_c{\mu}}{T}\Bigr]
	\right]
\ ,\\
X_{N_c}(m_q)
&=
\frac{\sinh[(N_c+1)E_q(m_q)/T]}{\sinh[E_q(m_q)/T]}\ ,
\label{Eq:FeffMq}
\end{align}
where we have regarded 
the inverse of the temporal extension
$1/N_\tau$ as temperature $T$,
and $E_q(m_q)=\mathrm{arcsinh}\,(m_q)$ represents 
the one dimensional quark excitation energy 
coming from the constituent quark mass $m_q=b_{\sigma}\sigma+m_0$.
%...] SCL ---------------------------------------------------------------------%

%%%%%%%%%%%%%%%%%%%%%%%%%%%%%%%%%%%%%%%%%%%%%%%%%%%%%%%%%%%%%%%%%%%%%%%%
%%%%%%%%%%%%%%%%%%%%%%%%%%%%%%%%%%%%%%%%%%%%%%%%%%%%%%%%%%%%%%%%%%%%%%%%

%...[ 1/g^2: Effective Action -------------------------------------------------%
We shall now evaluate NLO ($1/g^2$) correction terms
coming from the plaquette action in $S_\mathrm{LQCD}$.
We concentrate on the leading order of the $1/d$ expansion.
By integrating out spatial link variables,
two types of terms shown in the last two graphs in Fig.~\ref{Fig:plaq}
are found to appear from temporal and spatial plaquettes~\cite{Jolicoeur:1983tz,FP},
\begin{align}
&\Delta S_\beta^{(\tau)}
	=\frac{\beta_\tau}{4d}
		\sum_{x,j>0} 
		(
		 \Vp_x \Vm_{x+\hat{j}}
		+\Vp_x \Vm_{x-\hat{j}}
		)
\label{Eq:ActionGT}
\ ,\\
&\Delta S_\beta^{(s)}
	=\frac{-\beta_s}{d(d-1)}
		\sum_{x,0<k<j} 
		M_{x}
		M_{x+\hat{j}}
		M_{x+\hat{k}}
		M_{x+\hat{k}+\hat{j}}
	\ ,
\label{Eq:ActionGS}
\end{align}
where
$\beta_\tau=d/N_c^2g^2$
and 
$\beta_s=d(d-1)/8N_c^4g^2$.
%...] 1/g^2: Effective Action -------------------------------------------------%

%...[ Extended Hubbard Stratonovich transformation ----------------------------%
The NLO terms contain the product of different types,
such as $V^+V^-$ in $\Delta S_\beta^{(\tau)}$.
In order to treat these terms,
we propose a new mean field technique,
named Extended Hubbard-Stratonovich (EHS) transformation.
Let us consider to evaluate a quantity 
$e^{\alpha A B}$,
where $(A,B)$ and $\alpha$
represent arbitrary composite fields and an positive constant,
respectively.
We can represent $e^{\alpha AB}$ in the form of Gaussian integral
over two auxiliary fields $(\varphi,\phi)$,
\begin{align}
e^{\alpha A B}
&= \int\, d\varphi\, d\phi\,
	e^{-\alpha\left\{
		 (\varphi-(A+B)/2)^2
		+(\phi-i(A-B)/2)^2
		\right\}+\alpha A B}
\nonumber\\
&= \int\, d\varphi\, d\phi\,
	e^{-\alpha\left\{
		\varphi^2-(A+B)\varphi
		+ \phi^2 - i(A-B)\phi
		\right\}}
\label{Eq:EHSa}
\ .
\end{align}
The integral over the new fields $(\varphi,\phi)$
is approximated by the saddle point value,
$\varphi = \VEV{A+B}/2$ and $\phi = i\VEV{A-B}/2$.
Specifically 
in the case where both $\VEV{A}$ and $\VEV{B}$ are real,
which applies to the later discussion,
the stationary value of $\phi$ becomes pure imaginary.
Thus we replace $\phi \to i\omega$
and require the stationary condition for the real value of $\omega$,
\begin{eqnarray}
e^{\alpha A B}
&\approx& e^{-\alpha\left\{
		\varphi^2-(A+B)\varphi-\omega^2+(A-B)\omega
	\right\}}\Big|_{\mathrm{stationary}}
\label{Eq:EHSc}
\ .
\end{eqnarray}
In the case of $A=B$, Eq.~(\ref{Eq:EHSc}) reduces to
the standard HS transformation.
We find that the $e^{\alpha AB}$ is invariant
under the scale transformation, 
$A\to\lambda A$ and $B\to \lambda^{-1}B$.
This invariance is kept 
in rhs of Eq.~(\ref{Eq:EHSc}),
since the combinations
$\varphi-\omega=\langle A\rangle$
and
$\varphi+\omega=\langle B\rangle$
transform in the same way
as $A$ and $B$, respectively.
This means that the effective potential is independent
of the choice of $\lambda$.

We apply the EHS transformation to the NLO terms.
For the temporal and spatial plaquette action terms,
$\Delta S_\beta^{(\tau)}$ and $\Delta S_\beta^{(s)}$,
we substitute
$(\alpha,A,B)=(\beta_\tau/4d,-\Vp_x,\Vm_{x+\hat{j}})$
and 
$(\beta_s/d(d-1), M_xM_{x+\hat{j}},
M_{x+\hat{k}}M_{x+\hat{k}+\hat{j}})$ in Eq.~(\ref{Eq:EHSc}), 
respectively, and obtain,
\begin{align}
&\Delta S_\beta^{(\tau)}
\approx \frac{\beta_\tau}{4d}\sum_{x,j>0}\left[
		\varphi_\tau^2-\omega_\tau^2
	+(\varphi_\tau-\omega_\tau)\Vp_x(\mu)
		\right.
\left.
	-(\varphi_\tau+\omega_\tau)\Vm_x(\mu)
	\right]
	+(j\leftrightarrow -j)
\label{Eq:EHStauA}
\ ,\\
&\Delta S_\beta^{(s)}
\approx\frac{\beta_s}{d(d-1)}
\sum_{x,0<k<j}
\bigl[
\varphi_s^2-\omega_s^2
-(\varphi_s-\omega_s)M_xM_{x+\hat{j}}
-(\varphi_s+\omega_s)M_{x+\hat{k}}M_{x+\hat{k}+\hat{j}}
\bigr]
\label{Eq:EHSsA}
\ .
\end{align}

We can absorb NLO terms, $\Delta S_\beta^{(\tau,s)}$
in Eqs.~(\ref{Eq:EHStauA}) and (\ref{Eq:EHSsA}),
in the coefficient modification 
of the SCL effective action, $S_\mathrm{SCL}$ in Eq.~(\ref{Eq:SCL_Action}).
We assume that auxiliary fields take constant and isotropic values,
which are independent from the space-time $x$ and spatial directions $j, k$.
In $\Delta S_\beta^{(s)}$,
effects of $\omega_s$ disappear for constant auxiliary fields,
and $\varphi_s$ modifies the coefficient of $MM$ terms
in $S_\mathrm{SCL}$ as $-b_\sigma/2d \to -\tilde{b}_\sigma/2d$,
where $\tilde{b}_\sigma=b_\sigma+2\beta_s\varphi_s$.
Combined with the temporal hopping term in $S_\mathrm{LQCD}$,
the coefficients of $V^{\pm}$ are found to be,
$Z_\mp/2$, where $Z_\pm =1+\beta_{\tau}(\phi_{\tau}\pm\omega_{\tau})$.
We rewrite these coefficients as
$Z_\pm = Z_\chi \exp(\pm\delta\mu)$,
then NLO effective action containing fermions is written as,
\begin{align}
S_\mathrm{NLO}^{(F)}
= \frac{Z_\chi}{2}\sum_x\left[V^+(\tilde{\mu})-V^-(\tilde{\mu})\right]+\sum_x m_0 M_x
-\frac{\tilde{b}_\sigma}{2d}\sum_{x,j>0}M_xM_{x+\hat{j}}
+{\cal O}\Bigl(\frac{1}{\sqrt{d}},\frac{1}{g^2}\Bigr)
\ ,
\label{Eq:NLO_Action}
\end{align}
where $\tilde{\mu}=\mu-\delta\mu=\mu-\log\sqrt{Z_+/Z_-}$
and $Z_\chi=\sqrt{Z_+Z_-}$
represent shifted chemical potential
and the wave function renormalization factor.

Now we can repeat the same procedure as the standard SCL prescription.
Bosonization of the four-fermi interaction in
$S_\mathrm{NLO}$
leads to the quark mass term $(\tilde{b}_\sigma\sigma+m_0)M$,
and the quark integral and the temporal link integral
are also analogous to the SCL case.
The effects of wave functional renormalization factor 
$Z_\chi$ would be a little bit non-trivial:
Let us define the temporal hopping matrix as
$\sum_{xy}\chibar_x\Kern_{xy}^{(\tau)}\chi_y\equiv\sum_x[V^+_x-V^-_x]/2$,
then the quark determinant in SCL reads
$\mathrm{Det}\bigl[\Kern^{(\tau)}+\mathbf{1}(b_{\sigma}\sigma+m_0)\bigr]$.
With NLO effects, this determinant is replaced with,
%
%The quark determinant in SCL,
%$\mathrm{Det}\bigl[\Kern^{(\tau)}+\mathbf{1}(b_{\sigma}\sigma+m_0)\bigr]$,
%is now replaced with,
\begin{align}
\mathrm{Det}
\Bigl[Z_\chi\Kern^{(\tau)}(\tilde{\mu})+\mathbf{1}(\tilde{b}_{\sigma}\sigma+m_0)\Bigr]
=\mathrm{Det}\bigl[Z_\chi\bigr]
\mathrm{Det}\Bigl[\Kern^{(\tau)}(\tilde{\mu})+\mathbf{1}(\tilde{b}_{\sigma}\sigma+m_0)/Z_\chi\Bigr]
\ ,\label{Eq:QDet_NLO}
\end{align}
where $\mathrm{Det}$ represents the determinant of the temporal
and the color matrix.
From the factor $\mathrm{Det}\bigl[Z_\chi\bigr]$
in Eq.~(\ref{Eq:QDet_NLO}),
the additional term $-N_c\log Z_\chi$ appears in the effective potential.
The second factor in the second line of Eq.~(\ref{Eq:QDet_NLO})
leads to $\Vq$ with modified quark mass and chemical potential.
Note that the quark mass is also modified by $Z_\chi$.
Finally the effective potential is found to be,
\begin{align}
\Feff{eff}=&\Feff{aux}+\Vq(\tilde{m}_q;\tilde{\mu},T)
\ ,\label{Eq:Feff}\\
\Feff{aux}
=&\frac{\tilde{b}_{\sigma}}{2}\sigma^2
	+\frac{\beta_\tau}{2}(\varphi_\tau^2-\omega_\tau^2)
	+\frac{\beta_s}{2}\varphi_s^2-N_c\log Z_\chi
\ ,\label{Eq:Faux}\\
\tilde{m}_q
=&(\tilde{b}_{\sigma}\sigma+m_0)/Z_\chi
\ .\label{Eq:mqNLO}
\end{align}
%From the NLO effective potential $\Feff{eff}$,
We find that the plaquette contributes to the effective potential
in the modification of the wave function renormalization factor $Z_\chi$,
the quark mass $\tilde{m}_q$,
and the shift of the effective potential $\tilde{\mu}$
in addition to some auxiliary field terms, $\Feff{aux}$.

Since the $1/g^4$ contributions have ambiguities in the present NLO treatment,
we here compare the results in several ways of truncation.
In the first treatment, 
abbreviated as NLO-A,
% we do not perform any additional approximations.
$\Feff{eff}$ in Eq.~(\ref{Eq:Feff}) is treated as it is,
and we do not invoke any further approximations.
In the second treatment (NLO-B),
%${\cal O}(1/g^4)$ contributions in $Z_\chi$ and $\tilde{\mu}$ are removed,
${\cal O}(1/g^4)$ contributions in $Z_\chi$ and $\tilde{\mu}$ are truncated,
and simplified as
$Z_\chi=1+\beta_\tau\varphi_\tau$
and
$\tilde{\mu}=\mu-\beta_\tau\omega_\tau$.
In this treatment, we find that $\varphi_\tau$ and $\omega_\tau$ couple to quarks
separately through $\tilde{m}_q$ and $\tilde{\mu}$ in $\Vq$, respectively.
In the third prescription (NLO-C),
we further truncate ${\cal O}(1/g^4)$ terms in 
$\tilde{m}_q$ and in $\log Z_\chi$.
It is also possible to expand $\Vq$ 
with respect to $\delta\mu=\mu-\tilde{\mu}$ (NLO-D),
\begin{align}
\Vq(\tilde{m}_q;\tilde{\mu},T)
\simeq \Vq(\tilde{m}_q;\mu,T)-\beta_\tau\omega_\tau{\partial\Vq}/{\partial\mu}
\ .
\end{align}

In each treatment NLO-A,B,C, and D,
we evaluate $\Feff{eff}$ under the stationary condition
with respect to the auxiliary fields,
$\Phi=\sigma,\varphi_s,\varphi_{\tau},\omega_{\tau}$,
\begin{align}
\frac{\partial\Feff{eff}}{\partial\Phi}
=
\frac{\partial\Feff{aux}}{\partial\Phi}
+\frac{\partial\Vq}{\partial\tilde{m}_q}
\frac{\partial\tilde{m}_q}{\partial\Phi}
+\frac{\partial\Vq}{\partial\tilde{\mu}}
\frac{\partial\tilde{\mu}}{\partial\Phi}=0
\label{Eq:Stationary}\ .
\end{align}
Note that $\Vq$ depends on the auxiliary fields
via the two dynamical variables $\tilde{m}_q$ and $\tilde{\mu}$.
Here we show the stationary conditions in NLO-A, as an example.
Substituting $\sigma$ for $\Phi$ in Eq.~(\ref{Eq:Stationary}),
we obtain the relation,
$\sigma = -(1/Z_{\chi})(\partial\Vq/\partial m_{q})$.
By utilizing this result,
the stationary condition for $\varphi_s$
%$\Phi=\varphi_s$ in Eq.~(\ref{Eq:Stationary})
leads to $\varphi_s=\sigma^2$.
We can solve the coupled equation for
the stationary conditions
of $\varphi_{\tau}$ and $\omega_{\tau}$ as,
\begin{align}
\varphi_\tau=\frac{2\varphi_0}{1+\sqrt{1+4\beta_\tau\varphi_0}}
\ ,\qquad
\omega_\tau=-\frac{\partial\Vq}{\partial\tilde{\mu}}\equiv\rho_q
\label{Eq:StatSW}
\ ,
\end{align}
where $\varphi_0=N_c-Z_{\chi}\tilde{m}_q+\beta_{\tau}\omega_{\tau}^2$.
The second equation in Eq.~(\ref{Eq:StatSW}) indicates
the auxiliary field $\omega_{\tau}$ is 
nothing but the quark number density $\rho_q$.
The equilibrium condition is determined self-consistently
by minimizing $\Feff{eff}$ in terms of $\sigma$
under the constraint $\omega_{\tau}=\rho_q(T,\mu;\sigma,\omega_\tau)$.
Stationary conditions in NLO-B, C and D are solved similarly.
In NLO-D, $\omega_\tau$ is explicitly obtained as a function of $\sigma$,
$\omega_\tau=-\partial\Vq(m_q(\sigma);\mu,T)/\partial\mu$,
where the rhs does not contain $\omega_\tau$.
In this meaning, the NLO-D gives a similar formulation
to that in the previous work~\cite{Bilic}.

%%%%%%%%%%%%%%%%%%%%%%%%%%%%%%%%%%%%%%%%%%%%%%%%%%%%%%%%%%%%%%%%%%%%%%%%
\begin{figure}
\centerline{~\Psfig{7.5cm}{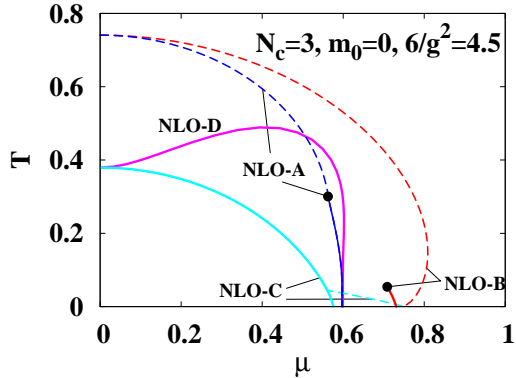}~}
\caption{
Phase diagram in NLO-A, B, C and D with $N_c=3, 6/g^2=4.5$
Solid and dashed lines show the first and second order phase transition
boundary, respectively.
}
\label{Fig:PD}
\end{figure}
%%%%%%%%%%%%%%%%%%%%%%%%%%%%%%%%%%%%%%%%%%%%%%%%%%%%%%%%%%%%%%%%%%%%%%%%

One of the most interesting features in the multi-order parameter
%(chiral condensate $\sigma$ and quark number density $\omega_\tau$)
treatments (NLO-A, B and C)
is that it predicts the existence of partially chiral restored (PCR) matter
at $N_c=3$ for relatively large $\beta=6/g^2$ values.
In Fig.~\ref{Fig:PD}, we show the phase diagram at $\beta=4.5$
in the chiral limit.
%First (second) order phase boundaries are shown in solid (dashed) curves.
In NLO-A and B,
the highest temperature of the first order phase boundary decreases,
and the critical point deviates from the second order phase transition boundary 
at $6/g^2 \simeq 4.5$ and $3.0$ in NLO-A and NLO-B, respectively.
In NLO-C, the second order critical chemical potential $\mu_c^{(2nd)}$ at $T=0$
overtakes the first order one at $6/g^2 \simeq 3.5$.
Between the first and second order phase boundaries, we find PCR matter.

%%%%%%%%%%%%%%%%%%%%%%%%%%%%%%%%%%%%%%%%%%%%%%%%%%%%%%%%%%%%%%%%%%%%%%%
\begin{figure}[bt]
\centerline{~\Psfig{7.5cm}{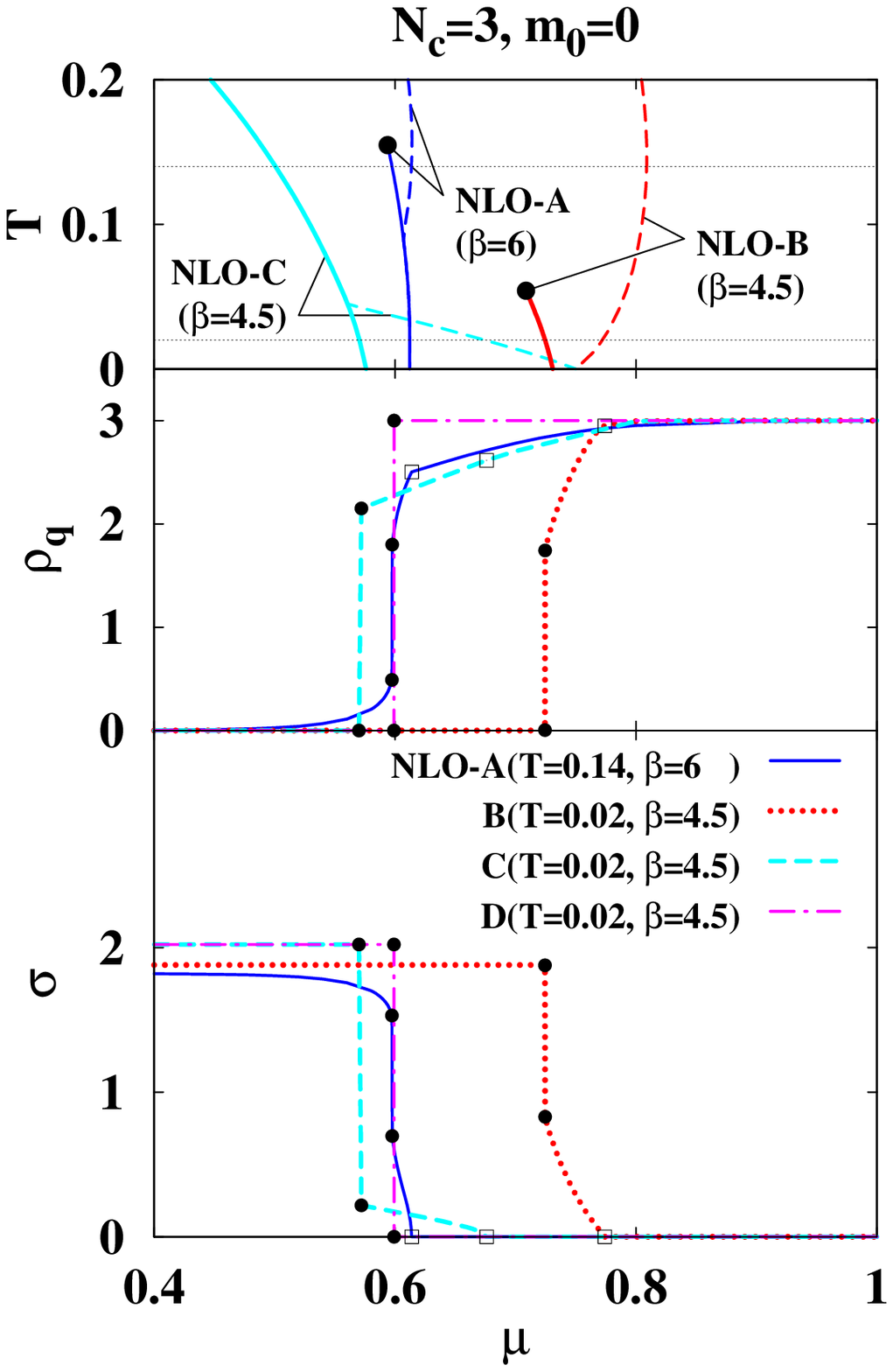}~}
\caption{
In the upper panel,
solid and dashed curves show the first and second order
phase transition boundaries,
and dots show the critical end point.
In the middle and lower panels,
solid, dotted, dashed, and dot-dashed curves
show the results in NLO-A, B, C and D, respectively,
and dots and open squares show the first and second order 
transition points.
}\label{Fig:pdrhosig}
\end{figure}
%%%%%%%%%%%%%%%%%%%%%%%%%%%%%%%%%%%%%%%%%%%%%%%%%%%%%%%%%%%%%%%%%%%%%%%

In Fig.~\ref{Fig:pdrhosig}, we show the comparison of $\rho_q$ and $\sigma$
in the present treatments NLO-A, B, C and D.
The gradual increase of the quark density is a common feature
of the multi-order parameter treatments.
At low temperatures,
we can investigate the appearance of the medium density matter
more intuitively.
% analytically.
The quark number density 
$\rho_q=-\partial\Feff{eff}/\partial\mu$ is evaluated as,
\begin{align}
\frac{\rho_q}{N_c}=\frac{2\sinh\bigl[N_c\tilde{\mu}/T\bigr]}
{X_{N_c}+2\cosh\bigl[N_c\tilde{\mu}/T\bigr]}
\ 
\raisebox{-2ex}{$\stackrel{\longrightarrow}{{\scriptstyle T\to 0}}$}
\ 
\frac{x^{N_c}}{1+x^{N_c}}
\ ,\label{Eq:rho}
\end{align}
where $x=\exp[-(E_q-\tilde{\mu})/T]$.
When $E_q>\tilde{\mu}$ is satisfied at small $T$,
we obtain $x\to 0$ and $\rho_q\to 0$,
while $E_q<\tilde{\mu}$ leads to $x\to\infty$ and $\rho_q\to N_c$.
%How can the medium density $0<\rho_q<N_c$ 
%be realized as NLO-B and NLO-C at small or zero temperature?
Medium density  $0<\rho_q<N_c$ can appear 
only in the case where the energy and chemical potential balances,
$E_q=\tilde{\mu}$, and 
%The condition $E_q=\tilde{\mu}$ could be the answer,
$x$ stays finite at $T=0$.
Since $\tilde{\mu}$ is a decreasing function of $\omega_\tau$,
we may have a medium density solution of Eq.~(\ref{Eq:StatSW}) in the region
$\tilde{\mu}(\sigma,\omega_\tau=N_c)<E_q(\sigma,\omega_\tau)<\mu$.
Specifically in NLO-B and C,
$E_q=\tilde{\mu}$ is found to be equivalent
to the density condition $\rho_q=(\mu-E_q)/\beta_{\tau}$,
which can take the a medium value.
In the large $\beta$ region,
this medium density matter can emerge in equilibrium
%gives the vacuum
as indicated in Fig.~\ref{Fig:pdrhosig}.
Also in NLO-A, medium density matter appears in a similar mechanism.
%similar mechanism would be dominant in the appearance of
Thus the multi-order parameter treatment 
is essential to obtain the medium density matter at low $T$,
and we observe chiral transitions twice as $\mu$ increases.
After the first one, the condition $E_q=\tilde{\mu}$
is approximately satisfied.
This state may correspond to the quarkyonic matter
where we expect $\mu\sim $(constituent quark mass)~\cite{McLerran:2008ua}.

In summary, we have evaluated the effective potential
in the next-to-leading order strong coupling lattice QCD (NLO SC-LQCD),
where $1/g^2$ effects are taken into account.
We have discussed the chiral transition 
and the possibility of the quarkyonic transition.
Here the order parameter of the latter is the quark number density,
which can be naturally introduced via the NLO effects.
The $\mathcal{O}(1/g^4)$ ambiguities have
been examined by comparing the results in several truncation schemes,
and we have found the following common properties
as far as the quark number density is treated as the order parameter
in addition to the chiral condensate.
(I)~The partially chiral restored (PCR) matter
can appear in the large $6/g^2$ region,
%when the quark number density is treated as the order parameter,
(II)~PCR sits next to the hadronic Nambu-Goldstone (NG) phase
in the $\mu$ direction,
(III)~the quark number density is high as $\Od(N_c)$ in PCR,
(IV)~after the ``NG$\to$PCR transition'',
the effective chemical potential is approximately the same as
the quark excitation energy,
(V)~and the chiral transition to the Wigner phase follows
after NG$\to$PCR transition.
All these properties would be the essence of
the quarkyonic matter and transition proposed in Ref.~\cite{QY}.
In the previous work, 
the quark-driven Polyakov loop evaluated in SC-LQCD is shown
to be small as ${\cal O}(1/N_c)$~\cite{FP},
and it would not grow much at low temperatures.
This feature also agrees with the proposed property of the quarkyonic matter.
In the present analyses,
we have found the two sequential chiral transitions
can occur along the $\mu$ direction at low $T$,
and the first one involves the quarkyonic transition.
Thus we can conclude that we have examined the quarkyonic picture 
with the clear connection to the finite coupling effects in the strong coupling expansion.
The detailed analyses of the phase diagram evolution
due to the finite coupling effects will be shown elsewhere~\cite{PDevol}.

There are many points to be improved in the present analysis.
First, ${\cal O}(1/g^4)$ ambiguities is not small, and
it is necessary to extend the analysis to the next-to-next-to-leading
order (NNLO).
Higher orders in $1/d$ expansion may be also necessary.
The quarkyonic transition
% condition $E_q=\tilde{\mu}$
may be related to the so-called baryon mass puzzle~\cite{Bringoltz,BMP};
$N_c\mu_c$ is calculated to be smaller than the baryon mass
in the strong coupling limit.
This means that the chiral transition takes place
before nuclear matter is formed.
The chemical potential shift discussed in this paper
may be the key to solve this problem.
Competition with the color superconducting (CSC) phase
and comparison with the MC results 
at finite baryon density~\cite{FKS,MC_LQCD}
are other interesting subjects to be investigated.

%\section*{Acknowledgements}
This work has been motivated by the discussions during
the international workshop on ``New Frontiers in QCD 2008''.
We would like to thank Prof. Larry McLerran, Prof. Kenji Fukushima,
and other participants in that workshop.
We would like to thank Prof.~Noboru Kawamoto
and Prof. Philippe de Forcrand for useful discussions.
This work was supported in part
by the Grant-in-Aid for Scientific Research from MEXT and JSPS
under the grant numbers,
    17070002		% (Tokutei, Strangeness)
and 19540252,		% ((C)(2), 2003), Ohnishi
the Global COE Program
"The Next Generation of Physics, Spun from Universality and Emergence",
and the Yukawa International Program for Quark-hadron Sciences (YIPQS).

%\appendix
%\section{First Appendix} %Empty argument \section{} yields `Appendix'. 
%
%\section{Second Appendix}

%%%%%%%%%%%%%%%%%%%%%%%%%%%%%%%%%%%%%%%%%%%%%%%%%%%%%%%%%%%%%
% Some macros are available for the bibliography:
%  o for general use
%    \JL : general journals                 \andvol : Vol (Year) Page
%  o for individual journal 
%    \AJ   : Astrophys. J.           \NC         : Nuovo Cim.
%    \ANN  : Ann. of Phys.           \NPA, \NPB  : Nucl. Phys. [A,B]
%    \CMP  : Commun. Math. Phys.     \PLA, \PLB  : Phys. Lett. [A,B]
%    \IJMP : Int. J. Mod. Phys.      \PRA - \PRE : Phys. Rev. [A-E]     
%    \JHEP : J. High Energy Phys.    \PRL        : Phys. Rev. Lett.
%    \JMP  : J. Math. Phys.          \PRP        : Phys. Rep.
%    \JP   : J. of Phys.             \PTP        : Prog. Theor. Phys.     
%    \JPSJ : J. Phys. Soc. Jpn.      \PTPS       : Prog. Theor. Phys. Suppl.
% Usage:
%  \PRD{45,1990,345}          ==> Phys.~Rev.\ \textbf{D45} (1990), 345
%  \JL{Nature,418,2002,123}   ==> Nature \textbf{418} (2002), 123
%  \andvol{B123,1995,1020}    ==> \textbf{B123} (1995), 1020
%%%%%%%%%%%%%%%%%%%%%%%%%%%%%%%%%%%%%%%%%%%%%%%%%%%%%%%%%%%%%

\end{document}